\documentclass[nofootinbib,twocolumn,showpacs,preprintnumbers,pre,aps,superscriptaddress]{revtex4-1}

\usepackage{graphicx}
\usepackage{bm}
\usepackage{amsmath}
\usepackage{amssymb}
\usepackage{color}

\begin{document}

\title{Non-reciprocal Phase Separations with Non-conserved Order Parameters}

\author{Maoji Liu}
\affiliation{Department of Physics, Wenzhou University, Wenzhou, Zhejiang 325035, China}
\affiliation{Wenzhou Institute, University of Chinese Academy of Sciences, 
Wenzhou, Zhejiang 325001, China}

\author{Zhanglin Hou}
\affiliation{Wenzhou Institute, University of Chinese Academy of Sciences, 
Wenzhou, Zhejiang 325001, China} 
\affiliation{Oujiang Laboratory, Wenzhou, Zhejiang 325000, China} 

\author{Hiroyuki Kitahata}
\affiliation{Department of Physics, Graduate School of Science, Chiba University, 
Chiba 263-8522, Japan} 

\author{Linli He}\email{Corresponding author: linlihe@wzu.edu.cn}
\affiliation{Department of Physics, Wenzhou University, Wenzhou, Zhejiang 325035, China}

\author{Shigeyuki Komura}\email{Corresponding author: komura@wiucas.ac.cn}
\affiliation{Wenzhou Institute, University of Chinese Academy of Sciences, 
Wenzhou, Zhejiang 325001, China} 
\affiliation{Oujiang Laboratory, Wenzhou, Zhejiang 325000, China} 
\affiliation{Department of Chemistry, Graduate School of Science, 
Tokyo Metropolitan University, Tokyo 192-0397, Japan} 


\begin{abstract}
We numerically investigate the phase separation dynamics of the non-reciprocal Allen-Cahn model 
in which two non-conserved order parameters are coupled. 
The system exhibits several dynamical patterns such as the randomly oscillating phase and the spiral 
phase as well as the homogeneously oscillating phase.
Topological defects in the spirals are either bound or unbound depending on the non-reciprocality.
The traveling stripe pattern is also found when the diffusion constants are highly asymmetric and 
the non-reciprocality is small.  
\end{abstract}

\maketitle


The concept of non-reciprocality has attracted considerable attention in the fields of active matter and 
non-equilibrium statistical mechanics~\cite{Ivlev2015,Loos20,Fruchart21}. 
When a non-equilibrium environment mediates the effective interactions among the constituent objects, 
the action-reaction symmetry (Newton's third law) can be violated. 
In such a situation, non-reciprocality leads to time-dependent structures in which spontaneously 
broken symmetries emerge dynamically~\cite{Fruchart21}.
For example, non-equilibrium chemical interactions between particles result in a new type of 
active phase separation~\cite{Agudo-Canalejo2019}.
Recently, non-reciprocality in stochastic active systems has been quantitatively characterized in terms 
of odd elasticity~\cite{Scheibner20,Fruchart22,Yasuda22,Lin23,Kobayashi23}.

Phase separation occurs commonly in natural phenomena and our daily lives, including 
multi-component mixtures~\cite{Doibook} and biological active systems~\cite{Hyman14,Berry18}.
The order parameters characterizing the phase separation are either non-conserved or 
conserved~\cite{Hohenberg77,ChaikinBook}, and the corresponding dynamics are described by the 
Allen-Cahn equation (Model A)~\cite{Allen1979} or the Cahn-Hilliard equation (Model B)~\cite{Cahn1958}.
Recently, several groups discussed the effects of non-reciprocal coupling between two phase-separating 
systems that have conserved order parameters, i.e.,
the non-reciprocal Cahn-Hilliard (NRCH) model~\cite{You2020,Saha2020,Tobias2022,Tobias2023,Suchanek2023,Brauns2023}.
It was shown that the non-reciprocality leads to the emergence of traveling patterns in purely diffusive 
systems, breaking both spatial and time-reversal symmetry.

Compared to the conserved case, however, the non-reciprocal Allen-Cahn (NRAC) model has not been 
studied in detail.
Although the corresponding ordinary differential equations were partially discussed in Ref.~\cite{Saha2020}, 
their pattern formation dynamics have not been studied systematically.
The NRAC model can be regarded as one type of reaction-diffusion model. 
With an increase in non-reciprocality, the local dynamics change from competitive dynamics to an 
activator-inhibitor one.
It is known that several activator-inhibitor models, such as the Keener-Tyson model~\cite{Tyson88} 
for Belousov-Zhabotinsky reaction~\cite{Cassani2021} and the FitzHugh-Nagumo 
model~\cite{FitzHugh61,Nagumo62}, exhibit a traveling wave.
In two-dimensional (2D) systems, such a traveling wave can form a spiral structure where a temporal 
oscillation can be seen everywhere except the phase singularity points, 
i.e., topological defects~\cite{StrogatzBook}.
Hence, it is important to investigate the defect dynamics of the NRAC 
model~\cite{Sugimura15,Tan2020,Liu21}.

In this Letter, we numerically investigate the phase separation and the pattern formation dynamics 
of the NRAC model in the regime of the activator-inhibitor-type local dynamics. 
As the non-reciprocality is increased, the system starts to oscillate in time and we find several 
dynamic patterns such as the randomly oscillating structure, the spiral patterns, and the homogeneously 
oscillating phase. 
For spiral structures, the clockwise and counterclockwise topological defects of the phase field can 
be either bound or unbound depending on the non-reciprocality.
The dynamical behaviors of the NRAC model are summarized in terms of the steady-state phase diagram.
We also find the traveling stripe pattern when the diffusion constants are highly asymmetric and the
non-reciprocality is small.


\begin{figure*}[tb]
\begin{center}
\includegraphics[scale=0.32]{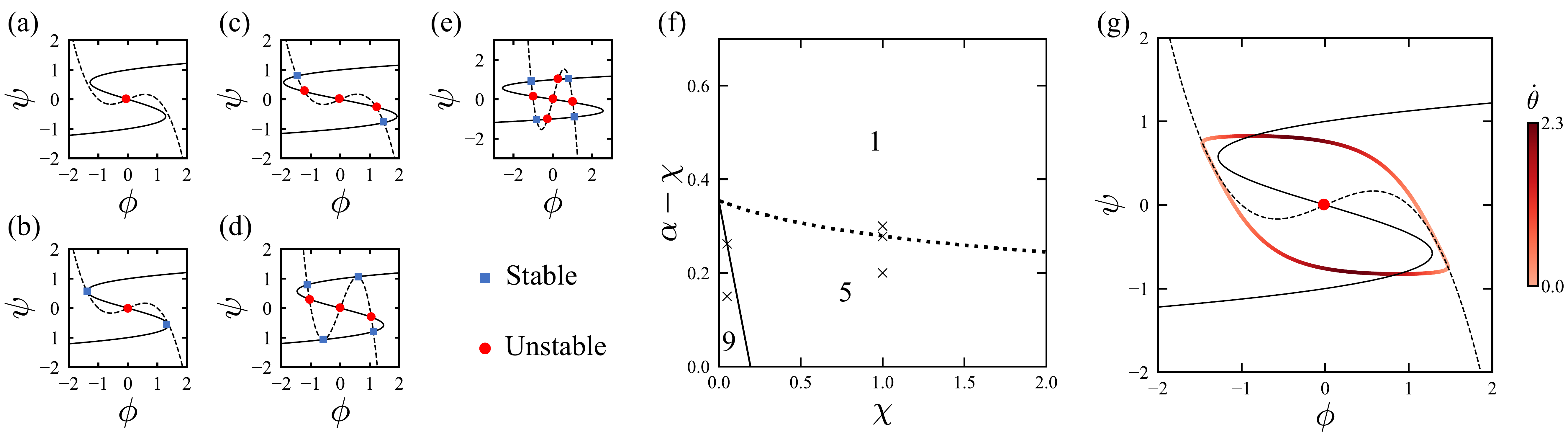}
\end{center}
\caption{
(Color online) Nullclines of Eqs.~(\ref{eq:modelphi}) and (\ref{eq:modelpsi}) corresponding to the 
solid ($\dot{\phi} = 0$) and dashed ($\dot{\psi} = 0$) lines, respectively.
The blue squares and red circles indicate the stable and unstable fixed points, respectively.  
The chosen parameters are 
(a) $(\chi, \alpha - \chi)=(1, 0.3)$, 
(b) $(1, 0.28)$,
(c) $(1, 0.2)$,
(d) $(0.05, 0.26)$, and 
(e) $(0.05, 0.15)$.
(f) The number of fixed points of nullclines.
The dotted and solid lines correspond to the cases of three and seven fixed points, respectively. 
The five crosses correspond to the parameters used in (a)-(e).
(g) The limit cycle when $(\chi, \alpha - \chi)=(1,0.3)$ for which there is only 
one unstable spiral at the origin. 
The solid and the dashed lines are the same nullclines shown in (a). 
The red color of the limit cycle indicates the angular velocity $\dot{\theta}$, i.e., the time derivative 
of the phase $\theta$ defined by Eq.~(\ref{eq:complex}).
}
\label{fig:nullcline}
\end{figure*}

Introducing 2D positional vector $\mathbf{r}$ and time $t$, we consider two phase-separating systems 
whose order parameters $\phi(\mathbf{r},t)$ and $\psi(\mathbf{r},t)$ are both non-conserved.
In the NRCA model, these two systems are coupled to each other in a non-reciprocal manner as given 
by the following coupled equations: 
\begin{align}
\dot{\phi} & =D_\phi \nabla^2 \phi +\phi-\phi^3-(\chi +\alpha )\psi,
\label{eq:modelphi}
\\
\dot{\psi} & =D_\psi \nabla^2\psi +\psi-\psi^3-(\chi -\alpha )\phi,
\label{eq:modelpsi}
\end{align}
where the dot indicates the time derivative.
In the above, $D_\phi$ and $D_\psi $ are the diffusion constants which are both positive, 
$\chi$ and $\alpha$ are the reciprocal and non-reciprocal coupling parameters, respectively,
between $\phi$ and $\psi$.
In this work, we assume $\chi \ge 0$ and $\alpha \ge 0$ without loss of generality 
as long as we allow the sign changes of $\phi$ and $\psi$.
When $\alpha=0$, both $\phi$ and $\psi$ will be homogeneously separated, as we explain later. 
The non-reciprocal coupling terms with the coefficient $\alpha$ cannot be derived from free energy 
and these terms are regarded as non-equilibrium chemical potentials~\cite{Saha2020}.

To discuss the instabilities in the system, we perform the linear stability analysis around 
the homogeneous state $(\phi,\psi)=(0,0)$. 
Let us introduce the Fourier components $\phi[\mathbf{k}, \omega]$ and $\psi[\mathbf{k}, \omega]$
such as by
\begin{align}
\phi(\mathbf{r},t)
=\int \frac{d\mathbf{k}}{(2\pi)^2} \frac{d \omega}{2\pi} \, \phi[\mathbf{k}, \omega]
\exp [i(\mathbf{k} \cdot \mathbf{r} - \omega t)],
\label{Fourier}
\end{align}   
where $\mathbf{k}$ is the 2D wavevector and $\omega$ is the frequency.
The linearized form of Eqs.~(\ref{eq:modelphi}) and (\ref{eq:modelpsi}) can be represented in the  
matrix form as 
\begin{equation}
-i \omega 
\begin{pmatrix}
\phi[\mathbf{k}, \omega] \\
\psi[\mathbf{k}, \omega]
\end{pmatrix}
=
\begin{pmatrix}
-D_{\phi} k^2 +1 & - (\chi+\alpha) \\
-(\chi-\alpha) & - D_{\psi} k^2 +1 
\end{pmatrix}
\begin{pmatrix}
\phi[\mathbf{k}, \omega] \\
\psi[\mathbf{k}, \omega]
\end{pmatrix},
\label{eq:per}
\end{equation}
where $k=\vert \mathbf{k} \vert$.
Then we obtain the following dispersion relation:
\begin{align}
-i\omega & = 1-\frac{D_\phi+D_\psi}{2} k^2
 \pm\sqrt{\chi^2-\alpha^2 + \frac{(D_\phi-D_\psi)^2}{4}k^4}.
\label{eq:eigenvalues}
\end{align}

In the long wavelength limit of $k \rightarrow 0$, the state $(\phi,\psi)=(0,0)$ is linearly unstable; 
it is a saddle point for $\chi > \sqrt{1 + \alpha^2}$ and an unstable node for $\alpha < \chi < \sqrt{1 + \alpha^2}$.
When $\alpha > \chi$, the right-hand side of Eq.~(\ref{eq:eigenvalues}) has an imaginary part with a positive real part.
This means that the state $(\phi,\psi)=(0,0)$ becomes an unstable spiral and the system oscillates in time.  
The oscillation frequency obtained within the linear stability analysis is given by 
$\Omega = \sqrt{\alpha^2-\chi^2}$.
Hereafter, we shall mainly focus on the case $\alpha \ge \chi$ and adopt $\alpha-\chi \ge 0$ 
as the quantity to control the degree of non-reciprocality.

\begin{figure*}[tb]
\begin{center}
\includegraphics[scale=0.32]{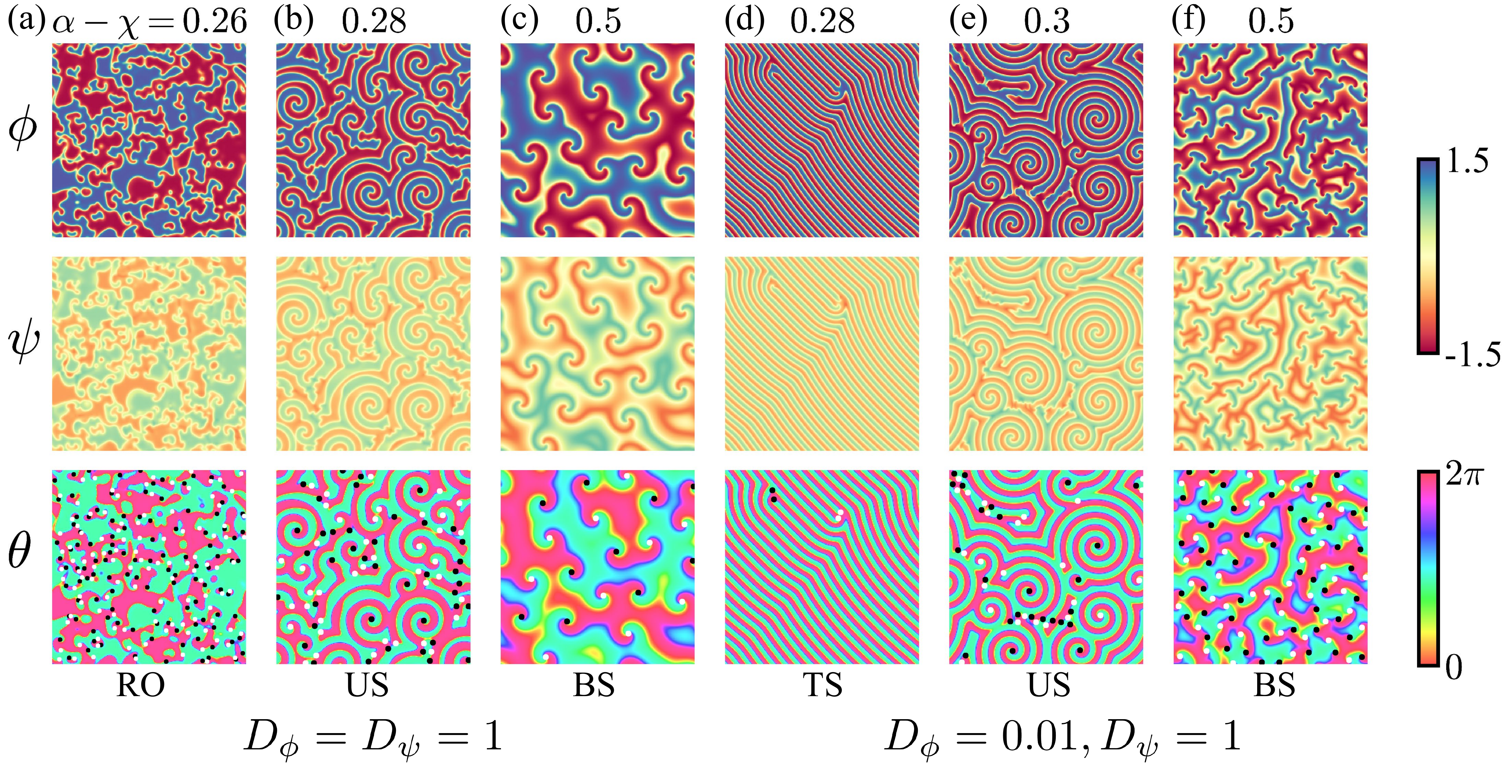}
\end{center}
\caption{
(Color online) The steady-state snapshot patterns of the order parameters $\phi$, $\psi$, and the phase 
$\theta$ [see Eq.~(\ref{eq:complex})] when $\chi=1$.
We choose the configuration data at $t=8\times10^4$.
(a)-(c) show symmetric diffusion cases, $D_\phi=D_\psi=1$, and 
(d)-(f) show antisymmetric diffusion case, $D_{\phi}=0.01$ and $D_{\psi}=1$. 
(a) $\alpha - \chi=0.26$ corresponding to the \textit{randomly oscillating phase} (RO)\cite{ROmovie},
(b) $\alpha - \chi=0.28$ corresponding to the \textit{unbound spiral phase} (US)\cite{USmovie}, 
(c) $\alpha - \chi=0.5$ corresponding to the \textit{bound spiral phase} (BS)\cite{BSmovie},
(d) $\alpha - \chi=0.28$ corresponding to the \textit{traveling stripe phase} (TS)\cite{TSmovie},
(e) $\alpha - \chi=0.3$ corresponding to the \textit{unbound spiral phase} (US), and 
(f) $\alpha - \chi=0.5$ corresponding to the \textit{bound spiral phase} (BS).
The black and white circles in the $\theta$-patterns represent $+1$ and $-1$ topological defects, 
respectively.
}
\label{fig:pattern}
\end{figure*}

Next, we discuss the nullclines of Eqs.~(\ref{eq:modelphi}) and (\ref{eq:modelpsi}) by setting 
$\dot{\phi}=\dot{\psi}=0$ in the absence of the diffusion terms (i.e., $k \to 0$)~\cite{StrogatzBook}. 
In Figs.~\ref{fig:nullcline}(a)-(e), we plot the nullclines for various sets of parameters showing different 
numbers of the fixed points.
The stability of the fixed point at $(\phi^\ast,\psi^\ast)$ can be determined by the eigenvalues of 
the following Jacobian matrix: 
\begin{equation}
\mathbf{J}=
\begin{pmatrix}
1-3 (\phi^\ast)^2 & -(\chi+\alpha) \\ 
-(\chi-\alpha) & 1-3 (\psi^\ast)^2 
\end{pmatrix}.
\end{equation}
In Figs.~\ref{fig:nullcline}(a)-(e), the stable and unstable fixed points are marked by the blue squares 
and red circles, respectively.
As mentioned above, the fixed point at $(\phi,\psi)=(0,0)$ is an unstable spiral. 
When there are five fixed points, as in Fig.~\ref{fig:nullcline}(c), two of the other four fixed points are stable 
nodes and the other two are saddle points.
When there are nine fixed points, as in Fig.~\ref{fig:nullcline}(e), four of the other eight fixed points are 
stable nodes and the other four are saddle points.
Figs.~\ref{fig:nullcline}(b) and (d) correspond to the cases when there are three and seven fixed points, respectively, 
in which pairs of fixed points are merged with saddle-node bifurcation.

The number of the fixed points is summarized in Fig.~\ref{fig:nullcline}(f).
The dotted and solid lines correspond to the cases of three and seven fixed points, respectively. 
The number of fixed points increases when $\chi$ and $\alpha-\chi$ are small.
Above the dotted line, on the other hand, there is only one unstable fixed point at $(\phi,\psi)=(0,0)$
and the system exhibits a limit cycle.
The structure of the limit cycle is independently presented in Fig.~\ref{fig:nullcline}(g) by using 
the same parameters as in Fig.~\ref{fig:nullcline}(a).


Next, we numerically solve Eqs.~(\ref{eq:modelphi}) and (\ref{eq:modelpsi}) by using the standard 
Euler's method on a 2D square lattice of size $256 \times 256$ with a periodic boundary condition. 
The time increment is taken to be $dt=0.1$ to explore long-time behaviors.
The initial values of the order parameters are chosen as $\phi_0=\psi_0=0$. 
Then, we add Gaussian random numbers with zero mean and variance $0.1$ to each simulation 
lattice as initial random configurations.
To determine the steady-state structure for a given set of parameters, we perform several 
simulation runs starting from different initial configurations.
Notice that the steady-state structures also include the dynamically oscillating patterns.
To discuss the dynamics of $\phi$ and $\psi$, it is convenient to introduce the following complex number 
\begin{align}
\phi(\mathbf{r},t)+i \psi(\mathbf{r},t) = \rho(\mathbf{r},t) e^{i\theta(\mathbf{r},t)},
\label{eq:complex}
\end{align}
where $\rho$ is the amplitude and $\theta$ is the phase.

\begin{figure*}[tb]
\begin{center}
\includegraphics[scale=0.32]{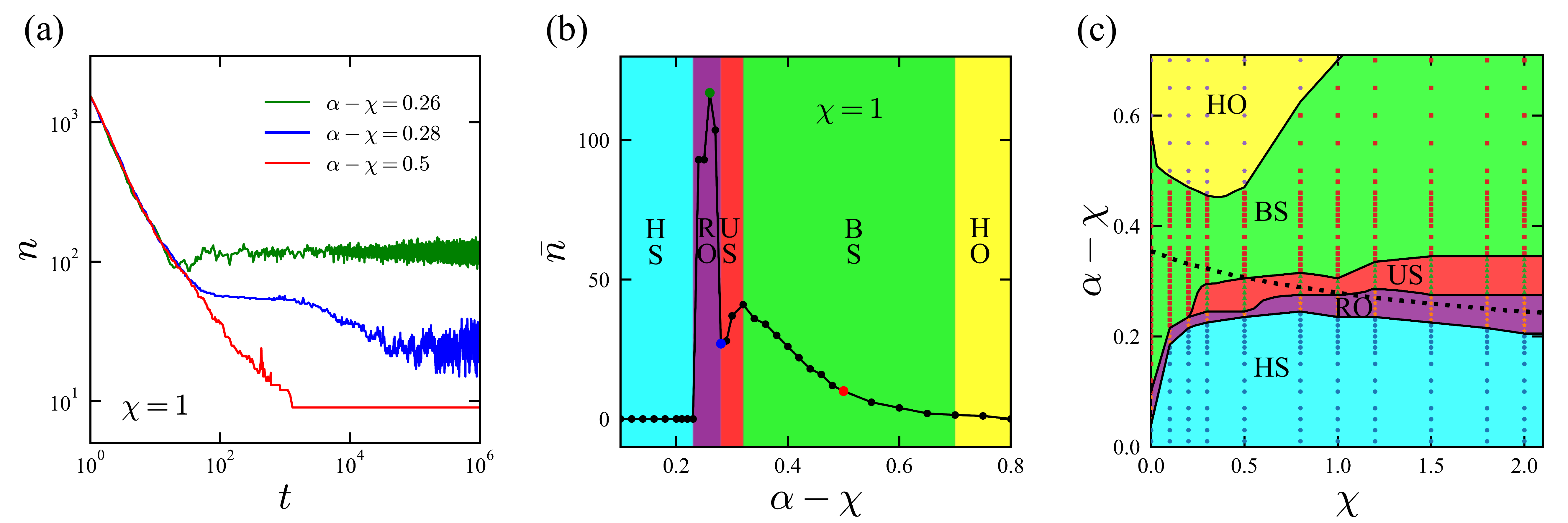}
\end{center}
\caption{
(Color online) (a) Time evolution of the topological defect density $n$ when $\chi=1$ and 
$\alpha - \chi=0.26$ (green, top), $0.28$ (blue, middle), and $0.5$ (red, bottom). 
(b) The average steady-state topological defect density $\bar{n}$ as a function of $\alpha - \chi$. 
We distinguish the following phases;
\textit{homogeneously separated phase} (HS),
\textit{randomly oscillating phase} (RO),
\textit{unbound spiral phase} (US),
\textit{bound spiral phase} (BS), and 
\textit{homogeneously oscillating phase} (HO).
The range of the error bar is smaller than the size of the symbol.
(c) The steady-state phase diagram of the NRAC model when $D_\phi = D_\psi=1$.
We also plot here the dotted line in Fig.~\ref{fig:nullcline}(f) above which the system exhibits a limit cycle.
}
\label{fig:defects}
\end{figure*}

We first discuss the simulation results when the diffusion constants are symmetric, i.e., $D_\phi= D_\psi=1$.
Fixing the parameter to $\chi=1$, we show in Fig.~\ref{fig:pattern} the sequence of the steady-state patterns 
as $\alpha-\chi$ is increased.  
We find five different steady states; 
(i) \textit{homogeneously separated phase} (HS), where $\phi$ and $\psi$ are spatially uniform at either one 
of the two stable states and they do not evolve in time, 
(ii) \textit{randomly oscillating phase} (RO), where each lattice site oscillates in time without forming any distinct structure
(see Fig.~\ref{fig:pattern}(a) and Ref.~\cite{ROmovie}), 
(iii) \textit{unbound spiral phase} (US), in which counterclockwise and clockwise spirals are formed 
while they are distributed randomly (see Fig.~\ref{fig:pattern}(b) and Ref.~\cite{USmovie}),  
(iv) \textit{bound spiral phase} (BS), in which counterclockwise and clockwise spirals form bound pairs
(see Fig.~\ref{fig:pattern}(c) and Ref.~\cite{BSmovie}), and 
(v) \textit{homogeneously oscillating phase} (HO), where the spatially uniform states oscillate 
in time, obeying the limit cycle behavior in Fig.~\ref{fig:nullcline}(g). 
For the RO, US, and BS phases, however, the distinction between them cannot be easily made only by the 
appearance of the dynamical patterns.

To quantitatively characterize the steady-state patterns including the spiral structures, 
we calculate the number of topological defects, $n$, per lattice site. 
Topological defects are singularity points where neighboring phase values $\theta(\mathbf{r},t)$ jump 
discontinuously.
The phase field defects have winding numbers of $+1$ or $-1$, corresponding to counterclockwise or 
clockwise rotating phase field structures.
Importantly, the numbers of $+1$ or $-1$ defects are identical because the total topological charges cancel 
each other, and we count only one of them.
The black and white circles in the $\theta$ patterns mark the numerically detected topological defects.

In Fig.~\ref{fig:defects}(a), we plot the time evolution of $n$ for different $\alpha-\chi$ values when $\chi=1$.
We see that $n$ initially decreases with time and eventually reaches a steady-state value 
(after about $t=10^4$) although it fluctuates around the average value.
In Fig.~\ref{fig:defects}(b), we plot the steady-state average value $\bar{n}$ as a function of $\alpha-\chi$.
Here, the average is taken over time after the system has reached a steady state and also 
over several different runs starting with different random initial configurations.
In the HS and HO phases, there are no topological defects because they are both homogeneous, while 
$\bar{n}$ takes large values in the RO phase. 
In the US phase, $\bar{n}$ drops dramatically compared to that of the RO phase but slightly increases 
as $\alpha-\chi$ is made larger. 
In the BS phase, however, $\bar{n}$ decreases monotonically with $\alpha-\chi$ until the system reaches 
the HO phase.
In Fig.~\ref{fig:defects}(b), we have used different colors to represent the different phases and most 
of the phase boundaries have been determined accurately. 
However, the boundary between the BS and HO phases depends on random initial configurations.
Hence, the BS-HO boundary was chosen as the minimum $\alpha-\chi$ value for which the HO phase 
appears at least in one simulation run among ten runs.
A more detailed analysis on the effects of initial configuration will be reported in our future work.

By repeating the same analysis for different $\chi$-values, we provide in Fig.~\ref{fig:defects}(c) the overall 
phase diagram when $D_\phi = D_\psi=1$. 
The sequence  
HS $\rightarrow$ RO $\rightarrow$ US $\rightarrow$ BS $\rightarrow$ HO can be generally seen 
for other $\chi$ values as the non-reciprocality, $\alpha-\chi$, is increased. 
The phase border between the HS and RO phases almost coincides with the dotted line in 
Fig.~\ref{fig:nullcline}(f) above which the system exhibits a limit cycle without the diffusion terms. 
The deviation between the dotted line and the phase border lines for small $\chi$ is purely due to 
the diffusion coupling in the NRAC model.
The phase border lines are weakly dependent on $\chi$, except the line separating between the 
BS and HO phases. 
The region of the BS phase becomes apparently larger as $\chi$ is increased. 
Although not shown in Fig.~\ref{fig:defects}(c), the HS phase extends to the region of 
$\alpha-\chi<0$ and $\alpha>0$. 
For $\alpha-\chi<0$ and $\alpha<0$, the same phase sequence 
(HS $\rightarrow$ RO $\rightarrow$ US $\rightarrow$ BS $\rightarrow$ HO) appears  
when $\vert \alpha-\chi \vert$ is increased.
For $\chi<0$, on the other hand, all the points $(\chi, \alpha-\chi)$ in Fig.~\ref{fig:defects}(c) 
transform to $(-\chi, \alpha+\chi)$.

The phase behavior in Fig.~\ref{fig:defects}(c) can partially be explained as follows. 
For the parameter regions where the local dynamics have multiple fixed points, the system can eventually 
converge to the homogeneous state at one of the stable fixed points. 
On the the dotted lines in Figs.~\ref{fig:nullcline}(b) and \ref{fig:defects}(c), the two sets of stable and 
unstable fixed points appear by saddle-node bifurcation. 
Such behavior is similar to the transition between the oscillatory and excitable states in the 
Keener-Tyson model~\cite{Tyson88,Cassani2021} and the FitzHugh-Nagumo model~\cite{FitzHugh61,Nagumo62}. 
Similar to these models, the spatio-temporal patterns with topological defects also appear in the NRAC model
even when the system has stable fixed points. 
This is because the boundary between RO and HS states is located just below the dotted line 
in Fig.~\ref{fig:defects}(c).

As shown in Fig.~\ref{fig:nullcline}(g), the red color on the limit cycle indicates the velocity of the phase, 
$\dot{\theta}$, and it becomes small for two separate blue regions.
Compared to the Keener-Tyson model and the FitzHugh-Nagumo model, 
the characteristic feature of the NRAC model is that $\dot{\theta}$ has the periodicity of $\pi$ rather than $2\pi$.
This is because Eqs.~(\ref{eq:modelphi}) and (\ref{eq:modelpsi}) are invariant under the simultaneous 
transformations $\phi \to -\phi$ and $\psi \to -\psi$.
Hence, the dynamics of $\theta$ can be approximately described by the following phase equation
\begin{align}
\dot{\theta} \approx \Omega - A \cos 2 (\theta-\theta_0),
\label{eq:phasedynamics}
\end{align}
where $A$ is the amplitude that satisfies $\Omega \ge A$ and $\theta_0$ is a constant phase. 
This  simplification is justified for the parameter sets close to the dotted line in 
Fig.~\ref{fig:defects}(c).
When $A$ is slightly less than $\Omega$, the phase $\theta$ takes a long time to pass through 
the bottlenecks~\cite{StrogatzBook}.  
From Eq.~(\ref{eq:phasedynamics}), the period of oscillation can be obtained as 
$T=2\pi/\sqrt{\Omega^2-A^2}$ that diverges when $A \rightarrow \Omega$. 
In contrast, when the parameter sets are far above the dotted line,
the difference of the angular velocity $\dot{\theta}$ does not play an important role. 
In such a case, the behaviors of the NRAC model are analogous to those described by the 
complex Ginzburg-Landau equation~\cite{Aranson02}.

Finally, we briefly discuss the case when the diffusion constants are highly asymmetric; $D_{\phi}=0.01$ 
and $D_{\psi}=1$.
The overall phase behavior is similar to that in Fig.~\ref{fig:defects}(c).  
One of the differences for the asymmetric diffusion case is the emergence of the \textit{traveling  
stripe phase} (TS), as shown in Fig.~\ref{fig:pattern}(d) and Ref.~\cite{TSmovie}, when both 
$\chi$ and $\alpha-\chi$ are small. 
In this case, an oriented stripe pattern moves with a finite velocity in the direction perpendicular 
to the orientation. 
Such a traveling pattern was also found in the NRCH model~\cite{You2020,Saha2020} and in chemically 
reacting mixtures~\cite{Okuzono01,Okuzono03,Luo23}.
Another feature of the asymmetric diffusion case is that the regions of the spiral phases (both US 
and BS) become larger.
As shown in Figs.~\ref{fig:pattern}(e) and (f), the formation of spiral structures can be seen more clearly.  
Details of the asymmetric diffusion case will be reported in our future publication 
including the perturbative analysis for bistable systems.


In summary, we have numerically investigated the phase separation dynamics and the pattern formation 
of the non-reciprocal Allen-Cahn model. 
As the non-reciprocality is changed, various dynamical phases emerge such as the randomly oscillating 
phase, unbound spiral phase, and bound spiral phase. 
The traveling stripe pattern can also be seen when the diffusion constants are highly asymmetric
and the non-reciprocality is small. 
In general, non-reciprocal phase separations are rich and useful because one can control the 
out-of-equilibrium behaviors of the intermediate patterns such as the traveling waves and spirals.
These dynamical intermediate structures cannot be observed in passively coupled phase 
separations without non-reciprocal interactions~\cite{Hirose09,Hirose12}.

Recently, we became aware of the work reporting that the NRCH model also admits 
stable spiral and target patterns~\cite{Rana23}. 
In their work, a transition from defect networks to traveling waves was also observed as the 
non-reciprocality is increased. 
In our work on the NRAC model, such a transition is observed only when the diffusion consents are 
highly asymmetric.


We thank S. Dobashi, T. Esashi, Ziluo Zhang, and Bin Zheng for useful discussion.
Z.H.\, L.H.\, and S.K.\ acknowledge the support by the National Natural Science Foundation of China 
(Nos.\ 12104453, 22273067, 12274098, and 12250710127) and the startup grant of Wenzhou Institute, 
University of Chinese Academy of Sciences (No.\ WIUCASQD2021041). 
H.K.\ acknowledges the support by the Japan Society for the Promotion of Science (JSPS) Core-to-Core 
Program ``Advanced core-to-core network for the physics of self-organizing active matter" (No.\ JPJSCCA20230002) 
and JSPS KAKENHI (No.\ JP21H01004).



\end{document}